\documentclass[twocolumn]{aastex701}
\usepackage{amsmath}
\usepackage{xcolor}

\newcommand{\mstar}{M$_{\star}$}
\newcommand{\msol}{M$_{\odot}$}
\newcommand{\logmass}{$\log(\mathrm{M}_\ast/\mathrm{M}_\odot)$}
\newcommand{\sfrone}{$\mathrm{SFR}_{100}$}
\newcommand{\logsfr}{$\log(\mathrm{SFR}_{100 \rm\,Myr}/[\mathrm{M}_\odot\mathrm{yr}^{-1}])$}
\newcommand{\ebvgas}{$E(B-V)_{\rm gas}$}
\newcommand{\ebvstar}{$E(B-V)_{\rm star}$}
\newcommand{\deltaebv}{$\Delta E(B-V)$}

\newcommand{\oiii}{[O\thinspace{\sc iii}]}
\newcommand{\neiii}{[Ne\thinspace{\sc iii}]}
\newcommand{\oii}{[O\thinspace{\sc ii}]}

\newcommand{\nii}{[N\thinspace{\sc ii}]}

\newcommand{\halpha}{H$\alpha$}
\newcommand{\hbeta}{H$\beta$}

\newcommand{\oiiilam}{[O\thinspace{\sc iii}]$\lambda5008$}
\newcommand{\oiilam}{[O\thinspace{\sc ii}]$\lambda\lambda3727,3730$}
\newcommand{\niilam}{[N\thinspace{\sc ii}]$\lambda6585$}

\begin{document}

\title{Balmer Decrements and Nebular-Stellar Reddening in JADES Galaxies at $2.7<z<7$}


\author[0009-0002-6186-0293]{Shreya Karthikeyan}
\affiliation{Department of Physics \& Astronomy, University of California, Los Angeles, 430 Portola Plaza, Los Angeles, CA 90095, USA}
\email[show]{skarthik@astro.ucla.edu}

\author[0000-0003-1249-6392]{Leonardo Clarke}
\affiliation{Department of Physics \& Astronomy, University of California, Los Angeles, 430 Portola Plaza, Los Angeles, CA 90095, USA}
\email{leoclarke@astro.ucla.edu}

\author[0000-0003-3509-4855]{Alice E. Shapley}
\affiliation{Department of Physics \& Astronomy, University of California, Los Angeles, 430 Portola Plaza, Los Angeles, CA 90095, USA}
\email{aes@astro.ucla.edu}

\author[0000-0001-9489-3791]{Natalie Lam}
\altaffiliation{NSF Graduate Research Fellow}
\affiliation{Department of Physics \& Astronomy, University of California, Los Angeles, 430 Portola Plaza, Los Angeles, CA 90095, USA}
\email{natalielamwy@astro.ucla.edu}

\author[0000-0003-4792-9119]{Ryan L. Sanders}
\affiliation{University of Kentucky, 506 Library Drive, Lexington, KY, 40506, USA}
\email{ryan.sanders@uky.edu}

\author[0000-0001-9687-4973]{Naveen A. Reddy}
\affiliation{Department of Physics \& Astronomy, University of California, Riverside, 900 University Avenue, Riverside, CA 92521, USA}
\email{naveenr@ucr.edu}

\author[0000-0001-8426-1141]{Michael W. Topping}
\affiliation{Steward Observatory, University of Arizona, 933 N Cherry Avenue, Tucson, AZ 85721, USA}
\email{michaeltopping@arizona.edu}

\author[0000-0003-2680-005X]{Gabriel B. Brammer}
\affiliation{Cosmic Dawn Center (DAWN), Denmark}
\affiliation{Niels Bohr Institute, University of Copenhagen, Jagtvej 128, DK-2200 Copenhagen N, Denmark}
\email{gabriel.brammer@nbi.ku.dk}

\begin{abstract}
We aim to characterize nebular and stellar reddening in star-forming galaxies as a function of global galaxy properties (stellar mass, SFR, metallicity) at $2.7 < z< 7.0$. We also provide a prescription to convert SED-based \ebvstar{} to \ebvgas{} when direct measurements of nebular reddening are unavailable. Our results are based on JWST/NIRSpec measurements of both individual spectra, with a sample of 283 galaxies, and composite spectra, including a larger sample of 327 galaxies. We estimate nebular reddening using the Balmer decrement (\halpha{}$/$\hbeta{}) above $10^{8.5}$ \msol{}, where the sample is representative. Stellar reddening and SFRs are derived through \textsc{Prospector} SED fitting, while gas-phase metallicities are estimated from strong emission-line ratios. At fixed stellar mass, Balmer decrements remain consistent within uncertainties across our redshift range, indicating that stellar mass primarily determines the overall dust column even by $z \sim 7$. We find that differential reddening (\deltaebv{} $\equiv$ \ebvgas{} $-$ \ebvstar{}) scales linearly with mass and SFR at $z \sim 2.7 - 4.0$, but shows no mass or SFR dependence above $z \sim 4.0$. We find evidence for smaller \deltaebv{} above $z \sim 5.0$, suggesting that nebular emission and stellar continuum probe increasingly similar dust columns towards higher redshift. Finally, we find that nebular reddening correlates strongly with metallicity out to $z \sim 5$, whereas the correlation between stellar reddening and metallicity is weaker or absent. Together, these results suggest that both dust mass and geometry play a significant role in shaping the observed reddening of high-redshift galaxies.
\end{abstract}


\section{Introduction} \label{sec:intro}

Interstellar dust, composed of solid grains enriched in metals, is produced largely during the final stages of stellar evolution. Despite accounting for only a small fraction of the baryonic mass in galaxies, dust greatly influences their spectral energy distributions (SEDs). By absorbing and scattering UV/optical light produced in galaxies and re-emitting it at far-infrared and submillimeter wavelengths, dust alters both the emission lines and the continuum observed in galaxy spectra. If not properly accounted for, dust can lead to biased estimates of stellar mass and star-formation rate (SFR). Characterizing how dust attenuates galaxy light is therefore essential for recovering intrinsic galaxy properties. Quantifying dust attenuation requires theoretically or empirically motivated models that connect the dust, stars, and gas in galaxies.

Dust attenuation can be estimated using many different observational tracers. Nebular reddening is commonly quantified using Balmer recombination lines, which provide a robust measure of wavelength-dependent dust attenuation along sightlines to H II regions. Attenuation of the stellar continuum is typically inferred from SED fitting and/or rest-frame UV/optical colors, which indicate the degree of reddening of a galaxy's integrated stellar light. Stellar reddening is generally more sensitive to diffuse dust affecting the spatially extended stellar populations that dominate the continuum. Determining how both nebular and stellar reddening scale with global galaxy properties (e.g., stellar mass, SFR, metallicity) over a range in redshift provides empirical constraints on dust content and geometry. 

In particular, nebular reddening exhibits a well-established correlation with stellar mass: more massive galaxies exhibit a greater degree of nebular reddening on average (e.g., \citealt{Dominguez+13}; \citealt{Kashino+13}; \citealt{Price+14}). Multiple ground- and space-based studies find that this relation shows no significant evolution out to z $\sim$ 2 (e.g., \citealt{Whitaker+17}; \citealt{McLure+18}; \citealt{Shapley+22}). Recent JWST results suggest that this apparent non-evolution may extend to even higher redshifts (i.e., $z \sim 7$; \citealt{Shapley+23}; \citealt{Woodrum+25}). This invariance means that dust models must reproduce similar reddening at fixed mass despite evolving ISM conditions, motivating tests of whether attenuation trends with other galaxy properties such as SFR and metallicity change at earlier times.

Metallicity is a particularly interesting quantity for interpreting attenuation trends because it traces the available gas-phase metals that can be incorporated into dust through accretion and grain growth in the ISM (\citealt{Popping+17}; \citealt{Li+19}). As galaxies become chemically enriched, the dust-to-gas and dust-to-metals ratios increase (e.g., \citealt{Remy+14}), boosting both the overall dust content and the strength of attenuation. Results from $z \sim 0$ out to $z \sim 2$ show a well-established positive correlation between dust reddening and gas-phase metallicity (e.g., \citealt{Heckman+98}; \citealt{Reddy+10}; \citealt{Theios+19}; \citealt{Shivaei+20}), consistent with this picture. Since ISM conditions at high redshift differ markedly from those in the local universe (e.g. \citealt{Reddy+23b}; \citealt{Topping+25}), metallicity provides a more direct tracer of the material from which dust is produced. Thus, examining attenuation-metallicity trends offers complementary information about dust production and growth beyond what is captured by mass alone. Examining the strength of metallicity trends with both nebular and stellar reddening can reveal the role of chemical enrichment in regulating dust columns along different sightlines in a galaxy.

A key related constraint on dust geometry comes from comparing the attenuation toward nebular regions versus the attenuation towards the stellar continuum, revealing how dust is spatially distributed relative to star-forming gas and the overall stellar population. In star-forming galaxies, nebular emission lines are typically more reddened than the stellar continuum (\citealt{Calzetti+00}; \citealt{Charlot+00}; \citealt{Price+14}; \citealt{Reddy+15}). This difference is consistent with a scenario in which OB associations and their surrounding nebular regions experience additional, localized obscuration from birth-cloud dust relative to the diffuse ISM dust affecting the more extended stellar population. Out to z $\sim$ 2, this nebular-stellar reddening relationship has been examined as a function of galaxy properties such as stellar mass, SFR, and dust geometry (\citealt{Reddy+15}; \citealt{Reddy+20}; \citealt{Lorenz+23}). These studies find that nebular reddening is less offset from stellar reddening at $z \sim 2$ than in the local universe, indicating evolution in dust distributions at higher redshift. In particular, \citet{Reddy+20} suggested that the offset may not be driven purely by the different stellar populations traced by nebular lines and the UV continuum, since the UV continuum is also dominated by young OB associations. Instead, the offset could also reflect variations in the dust covering fraction and geometry of H II regions. Initial strides have been made towards extending these comparisons to higher redshifts (\citealt{Tsujita+25}), but the picture regarding the extent and drivers of any potential redshift evolution is still incomplete as the current samples analyzed at $z \geq 3$ are small.

JWST/NIRSpec now enables Balmer-line measurements at $z \geq 3$ (\citealt{Shapley+23}; \citealt{Sandles+23}), and JWST/NIRSpec datasets at $z \geq 3$ allow for joint determinations of nebular and stellar reddening across a broad range of stellar mass and SFR. A practical need has emerged in the high-redshift regime. Many spectra, in particular those from the NIRCam grism or with a singular NIRSpec grating, contain only one strong Balmer line (e.g., \citealt{Matharu+24}; \citealt{Covelo+25}), with the remaining line fluxes inferred from photometry. For such cases, nebular reddening is not measured directly, and must be inferred indirectly based on the stellar reddening and an empirical relationship between nebular and stellar reddening based on SED properties. \citet{Sanders+21} have established such a relationship for $z \sim 2$ star-forming galaxies, but the relationship between nebular and stellar reddening has not yet been characterized for a large sample at high redshift. Such a relationship enables robust nebular dust corrections even when a Balmer decrement measurement is unavailable.

We investigate the nature of dust in galaxies at $2.7 \leq z \leq 7$ using data from the JWST Advanced Deep Extragalactic Survey (JADES; \citealt{Eisenstein+23}; \citealt{Bunker+24}), assembling a spectroscopic sample of 293 individual galaxies, and constructing composite spectra incorporating 327 galaxies. We examine how the Balmer decrement varies with stellar mass and redshift, model differential reddening (\ebvgas{}$-$\ebvstar{}) as a function of mass and SFR, and test how reddening depends on gas-phase metallicity inferred from strong rest-frame optical nebular emission-line ratios.

The outline of this paper is as follows. Section \ref{sec:methods} describes the parent JADES sample, line-flux measurements, SED fitting used to determine \ebvstar{} values, and metallicity calibration. We also describe the method for constructing composite spectra and measuring line fluxes from the composites. In Section \ref{sec:results}, we present our results on the Balmer decrement versus stellar mass relation, relative reddening of nebular emission versus stellar continuum, and attenuation trends with metallicity. We interpret these results in the context of galaxy evolution and current dust models in Section \ref{sec:discussion}, and conclude with a brief summary of our findings in Section \ref{sec:conclusion}. Throughout this work, we assume a $\Lambda$CDM cosmology with $H_0 = 70 \; \rm km \; s^{-1} \; Mpc^{-1}$, $\Omega_m = 0.3$, and $\Omega_{\Lambda} = 0.7$, and $Z_\odot = 0.014$ \citep{Asplund+21}.

\section{Observations and Measurements} \label{sec:methods}

\subsection{Photometric and Spectroscopic Data Analysis}
\subsubsection{JADES NIRCam and NIRSpec Data}

In this study, we use data from DR3 of the JADES GTO and GO programs (PIDs: 1180, 1181, 1210, 1286, 3215; \citealt{Rieke+23}; \citealt{Eisenstein+23}; \citealt{Bunker+24}; \citealt{D’Eugenio+25}), which cover the GOODS-N and GOODS-S deep fields. A comprehensive description of the data reduction and catalog construction is provided in \cite{Clarke+25}. Here we summarize only the aspects most relevant to this work. 

The photometric catalog includes JWST/NIRCam imaging in 10 filters from 0.70-4.4$\mu$m. The JADES imaging products include NIRCam data from FRESCO \citep{Oesch+23} and JEMS \citep{Williams+23}. We retrieved data products from the JADES, FRESCO, and JEMS High-Level Science Projects via the Mikulski Archive for Space Telescopes (MAST) at the Space Telescope Science Institute (STScI) \citep{https://doi.org/10.17909/8tdj-8n28,https://doi.org/10.17909/gdyc-7g80,https://doi.org/10.17909/fsc4-dt61}. Spectroscopic observations were obtained with the JWST/NIRSpec Micro-Shutter Assembly (MSA; \citealt{Ferruit+22}), using both the low-resolution ($R \sim 100$) prism and the medium-resolution ($\rm R\sim 1000$) gratings. The grating observations employed the G140M/F070LP, G235M/F170LP, and G395M/F290LP grating/filter combinations. For the G140M/F070LP configuration, the DAWN JWST Archive (DJA)\footnote[1]{https://dawn-cph.github.io/dja/} spectra were truncated at 1.25$\mu \rm m$ to avoid contamination from higher-order dispersion in the $\sim 1.25$-$1.8 \mu\rm m$ range. We discuss the impact of this truncation on Balmer-line measurements in Section \ref{subsubsec:neb}.

As the JADES NIRSpec/MSA design allows overlapping spectral traces (see \citealt{D’Eugenio+25}), many of the medium-resolution spectra are contaminated by dispersion overlap from neighboring shutters. The target placement on the MSA is configured to avoid this contamination in the prism observations. However, as the prism traces have a shorter extent along the NIRSpec detector, the prism data are much less affected by this type of contamination. As such, we used the prism spectra matched to the NIRCam photometry for flux calibration, and primarily use the medium-resolution grating spectra for deblending closely spaced emission lines, such as H$\alpha$ and [N\,\textsc{ii}].

\subsubsection{SED Fitting}
\label{subsubsec:sed}

To infer the stellar population properties of our galaxies, we performed joint SED fitting of the NIRSpec prism spectra and NIRCam photometry using the \textsc{Prospector} SED-fitting code (\citealt{Johnson+21}). We adopted a non-parametric star formation history (SFH) using a Student-t distribution as a continuity prior, dividing the SFH into eight bins across cosmic time (e.g., \citealt{Tacchella+22}). The two most recent time bins were fixed to span 3 Myr and 10 Myr, respectively, while the remaining six bins were evenly distributed logarithmically in time back to redshift $z = 20$. For the stellar population synthesis model, we utilized the Flexible Stellar Population Synthesis (FSPS) models (\citealt{Conroy+09}; \citealt{Conroy+10}) with the MILES spectral library (\citealt{Sanchez+06}) and MIST isochrones \citep{2011ApJS..192....3P,2013ApJS..208....4P,2015ApJS..220...15P,2016ApJS..222....8D,2016ApJ...823..102C}. Further details on the SED-fitting process can be found in \cite{Clarke+25}.

As in earlier work (\citealt{Clarke+24}; \citealt{Topping+25}), each galaxy was fit under two sets of assumptions for the dust attenuation law and stellar metallicity. In one set (“SMC+0.28 Z$_\odot$”), we assumed an SMC attenuation law (\citealt{Gordon+03}) and fixed the stellar metallicity to 0.28 Z$_\odot$. In the other (“Calz+1.4 Z$_\odot$”), we assumed a \cite{Calzetti+00} attenuation law and fixed the stellar metallicity to 1.4 Z$_\odot$. For each object, we adopted as our fiducial solution the \textsc{Prospector} fit with the higher output maximum probability.

Across all redshift bins, the large majority of our final sample is assigned the SMC+0.28 Z$_\odot$ solution as the fiducial fit. In the $2.7 < z \leq 4.0$, $4.0 < z \leq 5.0$, $5.0 < z \leq 6.0$, and $6.0 < z \leq 7.0$ bins, the numbers of galaxies assigned the Calz+1.4 Z$_\odot$ solution are 22 (11\%), 9 (11\%), 11 (28\%), and 3 (19\%), respectively, while the numbers assigned the SMC+0.28 Z$_\odot$ solution are 185 (89\%), 75 (89\%), 29 (72\%), and 13 (81\%), respectively.

\subsubsection{Emission Line Fitting}

Given the overlapping medium-resolution grating spectra and the higher-order dispersion affecting G140M/F070LP at 1.2–1.8 $\mu \rm m$, we adopted a joint fitting approach in which we simultaneously modeled the emission lines in both the prism and grating spectra. We fit [O\,\textsc{ii}]$\lambda \lambda$3727,3730, [Ne\,\textsc{iii}]$\lambda$3870, H$\gamma$, H$\beta$, [O\,\textsc{iii}]$\lambda \lambda$4960,5008, H$\alpha$, [N\,\textsc{ii}]$\lambda\lambda$6550,6585, and [S\,\textsc{ii}]$\lambda\lambda$6718,6733. Each emission line was modeled with a Gaussian line profile superimposed on the continuum, where the continuum was taken from the best-fit \textsc{Prospector} SED model. As mentioned previously, the flux-calibrated prism spectra set the absolute line flux scale, while the medium-resolution grating spectra were used to resolve closely spaced emission lines. In practice, this approach allowed the integrated line fluxes in the prism and grating spectra to differ, while requiring the line ratios to be consistent across both datasets during fitting. We additionally fit the Balmer emission-line fluxes using the best-fit SED from the stellar population modeling as the underlying continuum. This approach automatically accounts for stellar Balmer absorption. A full description of the emission-line fitting procedure is given in \cite{Clarke+25}.

\subsection{Spectroscopic Sample}
\label{subsec:sample}

The JADES parent sample comprises 4,086 objects (\citealt{D’Eugenio+25}), with 1,297 objects at $z \geq 1.4$. From this sample, we excluded targets with specific SFRs (sSFRs) less than $10^{-11} \, \mathrm{yr}^{-1}$ to remove quiescent galaxies and ensure emission line detections in our target spectra. Additionally, we excluded 297 galaxies with \halpha{} equivalent widths exceeding $10^6$ in the grating spectra, as these values are unphysical. We then required objects in our sample to have $>3 \sigma$ detections of H$\alpha$ and at least one other HI recombination line (either \hbeta{} or H$\gamma$), leaving 604 galaxies. We then excluded objects flagged as AGN due to broad Balmer-line components or \nii{}$/$H$\alpha$ flux ratios $> 0.5$. This AGN criterion required wavelength coverage of \nii{}, removing 24 objects. Additionally, truncation of the G140M/F070LP spectra at 1.25$\mu \rm m$ caused H$\beta$ (H$\gamma$) to fall outside the grating coverage for 97\% (73\%) of the objects at $1.4 \leq z \leq \rm 2.7$. As these galaxies lacked robust H$\beta$ or H$\gamma$ measurements, we restricted our sample to $z > \rm 2.7$, yielding 425 objects in total.

Because our final sample was subject to the same selection criteria as the \citet{Clarke+25} star-forming main-sequence sample, we refer to Section 2.4 of \citet{Clarke+25} for a comprehensive assessment of our sample completeness. We summarize the main points here. Below $10^{8.5} \, \rm M_\odot$, the spectroscopic sample is biased toward UV-bright galaxies, and thus missing lower-SFR systems at fixed stellar mass. Our sample is representative of the galaxy population down to $10^{8.5} \, \rm M_\odot$, containing 293 galaxies in this mass range.

As shown in Figure 2 of \citet{Clarke+25}, the median $M_{UV}$ of the spectroscopic sample at $10^{10} \, \rm M_\odot$ in the $4.0 < z \leq 5.0$ bin is significantly fainter than the \citet{Simmonds+24} best-fit trend for a larger mass-complete JADES sample. This anomalous trend suggests a similar lack of UV-bright targets in our $4.0 < z \leq 5.0$ subsample above $10^{10} \, \rm M_\odot$. The implications of this non-representativeness for our analysis are discussed in Section \ref{subsec:limitations}.

\begin{figure}[ht!]
    \centering
    \includegraphics[width=0.48\textwidth]{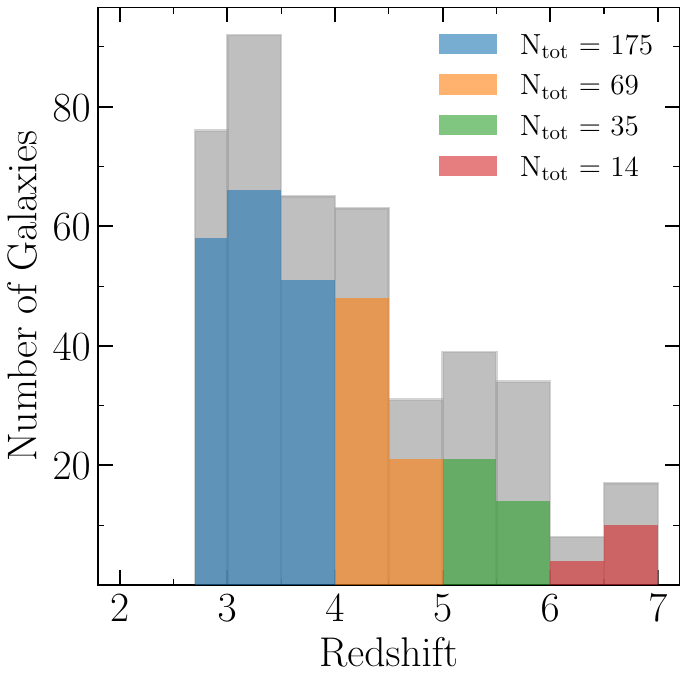}
    \caption{Galaxy counts by redshift for our sample. Grey bars indicate counts in the full observed mass range, while colored bars show counts in the mass-representative range (\logmass{} $\geq 8.5$). $N_{tot}$ denotes the total number of galaxies in each redshift bin in the mass-representative range.}
    \label{fig:1}
\end{figure}

Figure \ref{fig:1} shows the redshift distribution of our sample. Grey bars represent galaxies spanning the full observed mass range in each redshift bin, while color-coded bars (by redshift subsample) indicate counts in the mass-representative range only.

\subsection{Measurements}

\subsubsection{Nebular Reddening} \label{subsubsec:neb}

Our primary measurements of key nebular properties were derived from the Balmer decrement (\halpha{}/\hbeta{}), for which we required S/N $\geq 3$ in the H$\alpha$ and H$\beta$ line fluxes. We converted the Balmer decrements to the nebular reddening $E(B-V)_{\rm gas}$ assuming the average Milky Way extinction curve of \cite{Cardelli+89}. For 13 galaxies in our sample, H$\beta$ did not meet the S/N threshold but H$\gamma$ was detected at S/N $\geq$ 3; we instead used H$\alpha$/H$\gamma$ to calculate \ebvgas{} for these galaxies. For a given Balmer-line pair at wavelengths $\lambda_1$ and $\lambda_2$, we define $R_{\rm obs} = ({F_\lambda}_1/{F_\lambda}_2)_{\rm obs}$ and $R_{\rm int} = ({F_\lambda}_1/{F_\lambda}_2)_{\rm int}$, where $R_{\rm int}$ is the intrinsic Case B Balmer-line ratio. The corresponding nebular reddening is given by
\begin{equation}
E(B-V)_{\rm gas} = \frac{2.5}{k(\lambda_2)\,-\,k(\lambda_1)}\log_{10}\left( \frac{R_{\rm{obs}}}{R_{\rm int}}\right)
\end{equation}
where $k(\lambda)$ is the value of the \cite{Cardelli+89} curve at wavelength $\lambda$. For H$\alpha$/H$\beta$, we adopted $R_{\rm int} = 2.79$, and for H$\alpha$/H$\gamma$, we adopted $R_{\rm int} = 5.90$. These values correspond to standard Case B recombination under typical nebular conditions ($T_e = 15,000 \; \rm K$, $n_e =  10^2 \, \rm cm^{-3}$; \citealt{Reddy+23a}).

We estimated uncertainties on $E(B-V)_{\rm gas}$ by perturbing line fluxes 1000 times according to their measurement uncertainties and calculating a distribution of reddening values for each galaxy. As measurement noise can drive $R_{\rm obs} < R_{\rm int}$, which would correspond to unphysical negative $E(B-V)$ values, we set negative $E(B-V)_{\rm gas}$ values in each distribution to zero. We then adopted the median of the resulting distribution as our fiducial $E(B-V)_{\rm gas}$ and the 68\% interval as the $1\sigma$ uncertainty.

\subsubsection{Metallicity}
\label{subsubsec:met}

We estimated the gas-phase metallicity, $12 + \log(\mathrm{O/H})$, from the strong-line ratios [O\,\textsc{iii}]/H$\beta$, [O\,\textsc{iii}]/[O\,\textsc{ii}], and [Ne\,\textsc{iii}]/[O\,\textsc{ii}], using AURORA high-$z$ calibrations from \cite{Sanders+25}. For each ratio R, we evaluated a polynomial in $x\equiv 12 + \rm log(\mathrm{O/H})-8$ using the associated coefficients from Table 1 of \cite{Sanders+25}.

We fit $12 + \rm log(\mathrm{O/H})$ by minimizing the following expression:
\begin{equation}
\chi^2 =\sum_{i}\frac{\big(\log R_{i,{\rm obs}}-\log R_{i,{\rm cal}}(x)\big)^2}{\sigma_{i,{\rm obs}}^2+\sigma_{i,{\rm cal}}^2}
\end{equation}

where we summed over the set of line ratios used on a uniform metallicity grid $\in [7.3, 8.6]$ with step size $10^{-3}$ dex. $R_{i,{\rm obs}}$ indicates the observed value of the $i$th line ratio, and $R_{i,{\rm cal}}$ is the $i$th line ratio of the calibration at $x$. The $\sigma_{i,{\rm cal}}$ are the intrinsic scatters of the calibrations in $\log R$ at fixed $\mathrm{O/H}$ and are added in quadrature to the observed uncertainties, $\sigma_{i,{\rm obs}}$. We required at least one robustly detected ($> 3\sigma$) line ratio; furthermore, as [O\,\textsc{iii}]/H$\beta$ is not monotonic, we required at least one additional ratio when [O\,\textsc{iii}]/H$\beta$ was present. Uncertainties on metallicity were estimated from Monte-Carlo perturbation of the observed log-ratios. Specifically, we first determined the subset of ratios to be used in the best-fit $\chi^2$ solution, then perturbed each ratio in that subset 1000 times, re-minimizing $\chi^2$ on the same grid each time. We adopted the median of the resulting distribution as our $12 + \rm log(\mathrm{O/H})$ value, with the 68\% interval as the 1$\sigma$ uncertainty.

\begin{deluxetable*}{lccccc}
\tablecaption{Properties of Balmer Decrement Stacks \label{tab:balmer_stacks_props}}
\tablewidth{0pt}

\tablehead{
\colhead{$\log(M_\star/M_\odot)$} &
\colhead{$N$} &
\colhead{$\log(\mathrm{SFR_{100Myr}}/M_\odot\,\mathrm{yr}^{-1})$} &
\colhead{$\log(\mathrm{H}\alpha/\mathrm{H}\beta)$} &
\colhead{$E(B-V)_{\rm gas}$} &
\colhead{$E(B-V)_{\rm star}$}
}

\startdata
\multicolumn{6}{c}{$2.7 < z \leq 4.0$} \\
\tableline
8.64$^{+0.03}_{-0.03}$ & 
    40 & 
    -0.08 $\pm \; 0.06$ & 
    0.46$^{+0.02}_{-0.02}$ & 
    0.03$^{+0.05}_{-0.03}$ & 
    0.04 $\pm \; 0.01$ \\
8.89$^{+0.02}_{-0.02}$ & 
    40 & 
    0.07 $\pm \; 0.07$ & 
    0.46$^{+0.02}_{-0.02}$ & 
    0.04$^{+0.04}_{-0.04}$ & 
    0.04 $\pm \; 0.01$ \\
9.08$^{+0.02}_{-0.02}$ & 
    40 & 
    0.27 $\pm \; 0.07$ & 
    0.48$^{+0.02}_{-0.02}$ & 
    0.09$^{+0.05}_{-0.04}$ & 
    0.06 $\pm \; 0.02$ \\
9.35$^{+0.02}_{-0.02}$ & 
    39 & 
    0.57 $\pm \; 0.06$ & 
    0.51$^{+0.02}_{-0.01}$ & 
    0.15$^{+0.04}_{-0.03}$ & 
    0.07 $\pm \; 0.01$ \\
9.89$^{+0.03}_{-0.03}$ & 
    39 & 
    1.13 $\pm \; 0.11$ & 
    0.54$^{+0.01}_{-0.01}$ & 
    0.23$^{+0.02}_{-0.03}$ & 
    0.10 $\pm \;0.01$ \\
\tableline
\multicolumn{6}{c}{$4.0 < z \leq 5.0$} \\
\tableline
8.73$^{+0.03}_{-0.03}$ & 
    36 & 
    0.14 $\pm \; 0.05$ & 
    0.50$^{+0.02}_{-0.01}$ & 
    0.12$^{+0.04}_{-0.03}$ & 
    0.03 $\pm \; 0.01$ \\
9.36$^{+0.04}_{-0.05}$ & 
    36 & 
    0.67 $\pm \;0.9$ & 
    0.53$^{+0.01}_{-0.01}$ & 
    0.20$^{+0.03}_{-0.03}$ & 
    0.09 $\pm \; 0.02$ \\
\tableline
\multicolumn{6}{c}{$5.0 < z \leq 7.0$} \\
\tableline
8.74$^{+0.04}_{-0.05}$ & 
    29 & 
    0.37 $\pm \; 0.07$ & 
    0.49$^{+0.02}_{-0.02}$ & 
    0.10$^{+0.04}_{-0.04}$ & 
    0.05 $\pm \; 0.01$ \\
9.31$^{+0.04}_{-0.05}$ & 
    28 & 
    0.90 $\pm \;0.08$ & 
    0.51$^{+0.01}_{-0.01}$ & 
    0.15$^{+0.03}_{-0.03}$ & 
    0.10 $\pm \;0.01$ \\
\enddata

\end{deluxetable*}

\begin{deluxetable*}{lcccccc}
\tablecaption{Properties of Metallicity Stacks \label{tab:met_stacks_props}}
\tablewidth{0pt}

\tablehead{
\colhead{$\log(M_\star/M_\odot)$} &
\colhead{$N$} &
\colhead{$\log([\mathrm{O\,III}]/\mathrm{H}\beta)$} &
\colhead{$\log([\mathrm{O\,III}]/[\mathrm{O\,II}])$} &
\colhead{$\log([\mathrm{Ne\,III}]/[\mathrm{O\,II}])$} &
\colhead{$E(B-V)_{\rm gas}$} &
\colhead{$E(B-V)_{\rm star}$}
}

\startdata
\multicolumn{7}{c}{$2.7 < z \leq 4.0$} \\
\tableline
8.72$^{+0.04}_{-0.03}$ &
  42 &
  0.74$^{+0.02}_{-0.02}$ &
  0.61$^{+0.03}_{-0.03}$ &
  -0.47$^{+0.06}_{-0.07}$ &
  0.02$^{+0.04}_{-0.02}$ &
  0.03 $\pm \; 0.01$ \\
9.05$^{+0.02}_{-0.03}$ &
  42 &
  0.74$^{+0.02}_{-0.02}$ &
  0.47$^{+0.03}_{-0.03}$ &
  -0.59$^{+0.05}_{-0.05}$ &
  0.07$^{+0.04}_{-0.03}$ &
  0.06 $\pm \; 0.01$ \\
9.58$^{+0.05}_{-0.03}$ &
  42 &
  0.67$^{+0.01}_{-0.01}$ &
  0.28$^{+0.02}_{-0.02}$ &
  -0.83$^{+0.04}_{-0.04}$ &
  0.19$^{+0.03}_{-0.02}$ &
  0.09 $\pm \; 0.01$ \\
\tableline
\multicolumn{7}{c}{$4.0 < z \leq 5.0$} \\
\tableline
8.74$^{+0.03}_{2}$ &
  36 &
  0.76$^{+0.02}_{-0.02}$ &
  0.61$^{+0.03}_{-0.03}$ &
  -0.44$^{+0.03}_{-0.03}$ &
  0.12$^{+0.04}_{-0.03}$ &
  0.04 $\pm \; 0.01$ \\
9.34$^{+0.03}_{-0.05}$ &
  35 &
  0.72$^{+0.01}_{-0.01}$ &
  0.38$^{+0.02}_{-0.02}$ &
  -0.65$^{+0.03}_{-0.04}$ &
  0.20$^{+0.03}_{-0.03}$ &
  0.09 $\pm \; 0.02$ \\
\tableline
\multicolumn{7}{c}{$5.0 < z \leq 7.0$} \\
\tableline
8.73$^{+0.05}_{-0.05}$ &
  27 &
  0.76$^{+0.02}_{-0.02}$ &
  0.74$^{+0.04}_{-0.04}$ &
  -0.26$^{+0.04}_{-0.04}$ &
  0.10$^{+0.04}_{-0.04}$ &
  0.05 $\pm \; 0.01$ \\
9.29$^{+0.04}_{-0.05}$ &
  27 &
  0.76$^{+0.01}_{-0.01}$ &
  0.47$^{+0.02}_{-0.02}$ &
  -0.56$^{+0.03}_{-0.03}$ &
  0.14$^{+0.03}_{-0.03}$ &
  0.09 $\pm \; 0.01$ \\
\enddata

\end{deluxetable*} 

\subsection{Composite Spectra}

We present stacked composite spectra from the JADES sample, constructed to measure emission-line fluxes and galaxy properties at higher S/N. From the parent JADES sample, we excluded 297 galaxies with unphysical \halpha{} equivalent widths in the grating spectra, and those with sSFR $< 10^{-11} \, \mathrm{yr}^{-1}$ (one galaxy). Additionally, we required \halpha{} to be detected at $>3\sigma$ significance, excluding 353 objects. We then removed objects flagged as AGN due to broad Balmer-line components or \nii{}$/$\halpha{} flux ratios exceeding 0.5, discarding an additional 47 objects.

We generated two sets of stacks for this study that required wavelength coverage of certain emission lines to ensure that the line fluxes measured from composite spectra were representative of the galaxies contributing to each stack. One set required the individual component spectra to have wavelength coverage of the \halpha{} and \hbeta{} Balmer lines for the Balmer decrement analysis, and the other required wavelength coverage of \oii$\lambda\lambda 3727,3730$, \neiii$\lambda 3870$, H$\beta$, \oiii$\lambda 5008$, \nii$\lambda 6585$, and H$\alpha$ for the metallicity analysis. We refer to each of these sets of stacks as the “Balmer decrement stacks” and the “metallicity stacks,” respectively. After imposing these requirements, the Balmer decrement and metallicity stacks comprised 697 and 557 galaxies, respectively.

The galaxies were then divided into four redshift bins, namely $1.4\leq z < 2.7$, $2.7\leq z < 4.0$, $4.0\leq z < 5.0$, and $5.0\leq z < 7.0$. Hereafter, this work omits the $1.4 \leq z < 2.7$ redshift bin, as that range is not probed in our analysis. The Balmer decrement (metallicity) stacks included 488 (368) objects at $z \geq 2.7$. Within each redshift bin, we further divided galaxies into bins of equal number by stellar mass over the mass-representative range, and galaxies below $10^{8.5}\ M_{\odot}$ were stacked together in a single mass bin. In the mass-representative range, the Balmer decrement (metallicity) stacks incorporate a total of 327 (251) galaxies.

We first processed the grating spectrum for each galaxy by scaling the G140M, G235M, and G395M grating spectra to the flux calibrated prism spectrum. We calculated scaling factors between the line fluxes measured in the prism and grating spectra for the strong emission lines \oiilam, H$\gamma$, \hbeta{}, \oiiilam, \halpha{}, \niilam{}, Pa$\gamma$, Pa$\beta$, and Pa$\alpha$, requiring that each was detected at S/N $\geq3$. The grating-wide scaling factor was determined by the inverse-variance-weighted average of the emission lines that fall in the wavelength range of the grating. The three grating spectra were then stitched together, shifted into the rest frame, converted from flux density to luminosity density, and normalized to the non-dust-corrected \halpha{} luminosity measured from the prism spectrum. 

With the processed grating spectra, we then generated the composite spectra by re-sampling each spectrum onto a common wavelength grid using the \textsc{SpectRes} Python package \citep{Carnall+17}, with $\Delta \lambda$ as the median $\Delta \lambda$ of all galaxies in a given stack. We then calculated the median flux density at each wavelength. The above stacking process was repeated 100 times for each bin via Monte Carlo simulations to determine the uncertainties of the bin edges, median stellar mass, and median SFRs, as well as to generate the associated error spectrum. The 100 realizations of the composite spectrum for each bin were created by allowing galaxies to shift between mass bins according to the uncertainties on their stellar masses, bootstrap resampling within the bin to account for sample variance, and perturbing individual galaxy grating spectra by their error spectra prior to median-stacking. We then adopted the median flux and 1$\sigma$ standard deviation at each wavelength as the resultant composite spectrum and error spectrum. Emission-line fluxes were measured from single-Gaussian fits to the stacked spectra. The stacking process was also applied to the SED models associated with each galaxy to obtain a stacked continuum for the Balmer absorption correction.

For each stack, $E(B-V)_{\rm gas}$ was calculated from the measured Balmer decrement, adopting an intrinsic Case B recombination H$\alpha$/H$\beta$ ratio of 2.79, which assumes $T_e = 15,000 \; \rm K$ and $n_e =  10^2 \, \rm cm^{-3}$ (as noted in Section \ref{subsubsec:neb}). We estimated the associated \ebvstar{} and \logsfr{} (hereafter \sfrone{}) as follows. Here, \sfrone{} is the SFR averaged over the last 100 Myr, inferred from the \textsc{Prospector} SFH. For each stack, we bootstrap-resampled the quantity of interest from the individual galaxies contributing to the stack, perturbed each resampled value by its associated uncertainty, and calculated the median of the perturbed distribution. We repeated this procedure 1000 times. We then adopted the median and standard deviation of the resulting distribution of medians as the stack \ebvstar{} (or \sfrone{}) value and error, respectively. Metallicities were calculated using the procedure described in Section \ref{subsubsec:met}, using the measured line fluxes from the metallicity stacks. The resulting properties of the Balmer decrement and metallicity stacks are reported in Tables \ref{tab:balmer_stacks_props} and \ref{tab:met_stacks_props}, respectively.

\section{Results} \label{sec:results}

\subsection{Balmer Decrement versus Stellar Mass}

\begin{figure*}[ht!]
    \centering
    \includegraphics[width=0.95\linewidth]{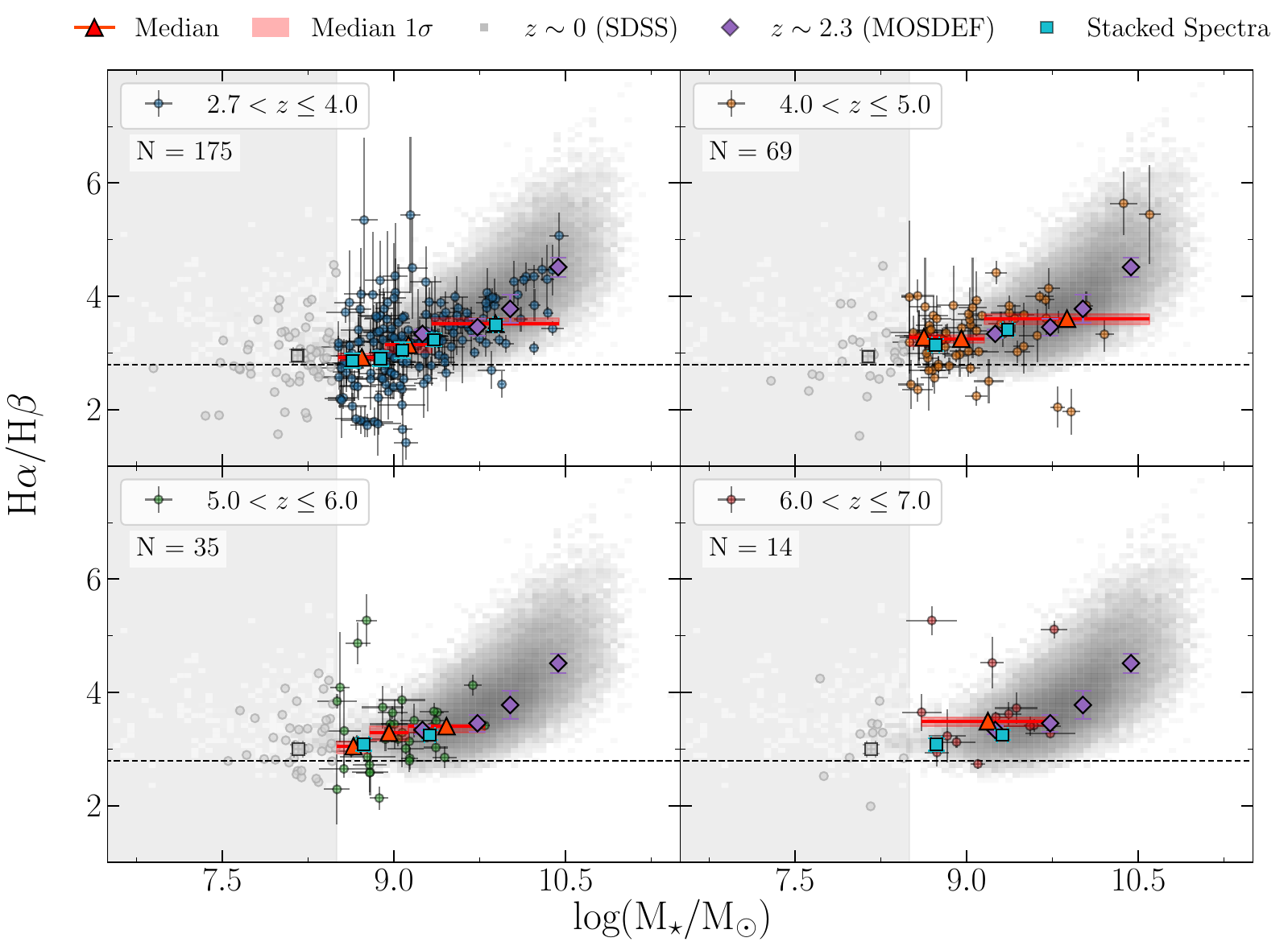}
    \caption{\halpha{}$/$\hbeta{} vs. stellar mass in bins of redshift. Binned medians of Balmer decrement are plotted as red triangles; bin widths are displayed as red line segments with the $1\sigma$ interval on \halpha{}$/$\hbeta{} shaded. The grey 2D histogram shows the $z \sim 0$ sample from SDSS; purple diamonds show $z \sim 2.3$ running medians in bins of \mstar{} from the MOSDEF survey \citep{Shapley+22}. Cyan squares indicate measurements from composite spectra. We show the combined $5.0 \leq z < 7.0$ composites in both the $5.0 < z \leq 6.0$ and $6.0 < z \leq 7.0$ redshift bins. The horizontal dashed line shows the assumed intrinsic Case B recombination value of 2.79. N denotes the number of galaxies in each redshift bin with \logmass{} $>$ 8.5. The greyed-out region and points on the left indicate measurements that are not in the mass-representative range of our sample.
    }
    \label{fig:2}
\end{figure*}

\begin{figure}[ht!]
    \centering
    \includegraphics[width=0.48\textwidth]{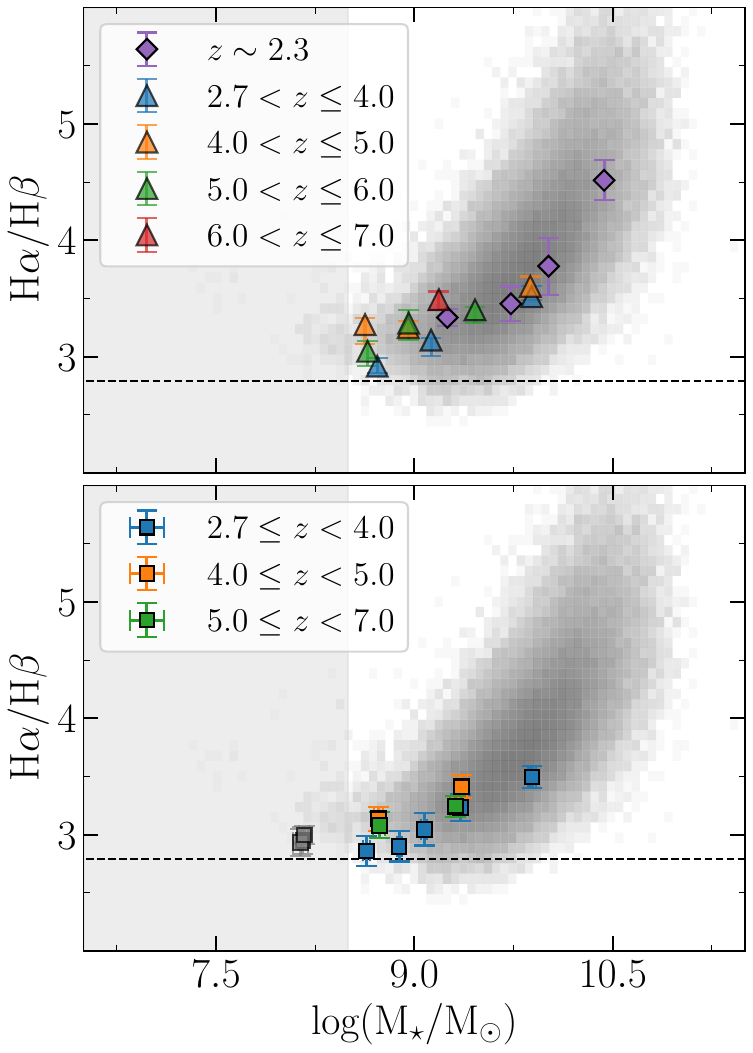}
    \caption{Comparison of Balmer decrement vs. stellar mass across redshift bins. The grey 2D histogram shows the $z \sim 0$ sample from SDSS. \textit{Top panel:} Binned medians from individual galaxies are plotted as triangles. $z \sim 2.3$ MOSDEF binned medians are shown as purple diamonds. \textit{Bottom panel:} Balmer decrement measurements from composite spectra.}
    \label{fig:3}
\end{figure}

We first investigate empirical trends between the observed Balmer decrement and $M_\star$. Figure \ref{fig:2} shows the Balmer decrement, H$\alpha$/H$\beta$, as a function of stellar mass for individual galaxies in our four redshift bins, along with the associated decrements from the composite spectra in cyan, with two comparison samples from \cite{Shapley+22}. We plot the comparison sample of z $\sim$ 0 galaxies from the Sloan Digital Sky Survey (SDSS; \citealt{Abazajian+09}) using a grey 2D histogram; running medians of H$\alpha$/H$\beta$ in bins of $M_\star$ from the MOSFIRE Deep Evolution Field (MOSDEF; \citealt{Kriek+15}) survey at z $\sim$ 2.3 are shown as purple diamonds. In the mass-representative range of our sample ($\rm log(M_\star/M_\odot) \geq 8.5$), we bin each redshift range in stellar mass with approximately equal numbers of galaxies per bin, and in each bin we compute the median Balmer decrement. The red points mark these medians, and the horizontal line segments span the corresponding mass-bin widths. Uncertainties on the medians are estimated via Monte Carlo simulations in which we perturb the line fluxes according to their measurement errors and recompute the medians. The resulting 16th-84th percentile range of the median is shown as the shaded region at each line segment. The cyan squares indicate the Balmer decrement measured from the composite spectra, which is based on a larger effective sample than the individual-galaxy measurements, since the stacks do not require an H$\beta$ detection. The horizontal dashed line marks the intrinsic Case B recombination ratio $R_{\rm int} = 2.79$. 

We find a statistically significant positive correlation between the Balmer decrement and stellar mass for individual galaxies in the two lowest redshift bins. We quantify the strength of the relationship using the Pearson correlation coefficient, $\rho$, and the associated statistical significance, $p$. For the $2.7 < z \leq 4.0$ bin, $\rho = 0.462$ ($p = 1.2 \times 10^{-10}$) and for $4.0 < z \leq 5.0$, $\rho = 0.40$ ($p= 6.54 \times 10^{-4}$). At $z \geq 5$, the correlation computed from the individual galaxies is not statistically significant, which may be due to selection effects from requiring individual H$\beta$ detections that potentially exclude the most dust-obscured systems from the sample. Additionally, given the limited number of individual galaxies in the $6.0 < z \leq 7.0$ bin, we have constructed only one mass bin and thus cannot test mass-dependence. Hence, we instead proceed to examine the stacked spectra in the combined $5.0 \leq z < 7.0$ bin, which incorporates an effectively larger sample due to alleviation of the \hbeta{} detection requirement.

To emphasize the mass-reddening relation and its redshift dependence, Figure \ref{fig:3} plots only the median and stacked H$\alpha$/H$\beta$ in each mass bin for all redshift bins on the same axes. At fixed stellar mass, the H$\alpha$/H$\beta$ ratios are consistent across redshift bins within the quoted uncertainties (typically $1\sigma$, and $\sim2\sigma$ at the lowest masses) for both median and stacked measurements, indicating little to no evolution in the mass-reddening relation out to z $\sim$ 7.

\subsection{Stellar-to-Nebular Reddening Conversion}
\label{subsec:conversion}

\begin{figure*}[ht!]
    \centering
    \includegraphics[width=0.95\linewidth]{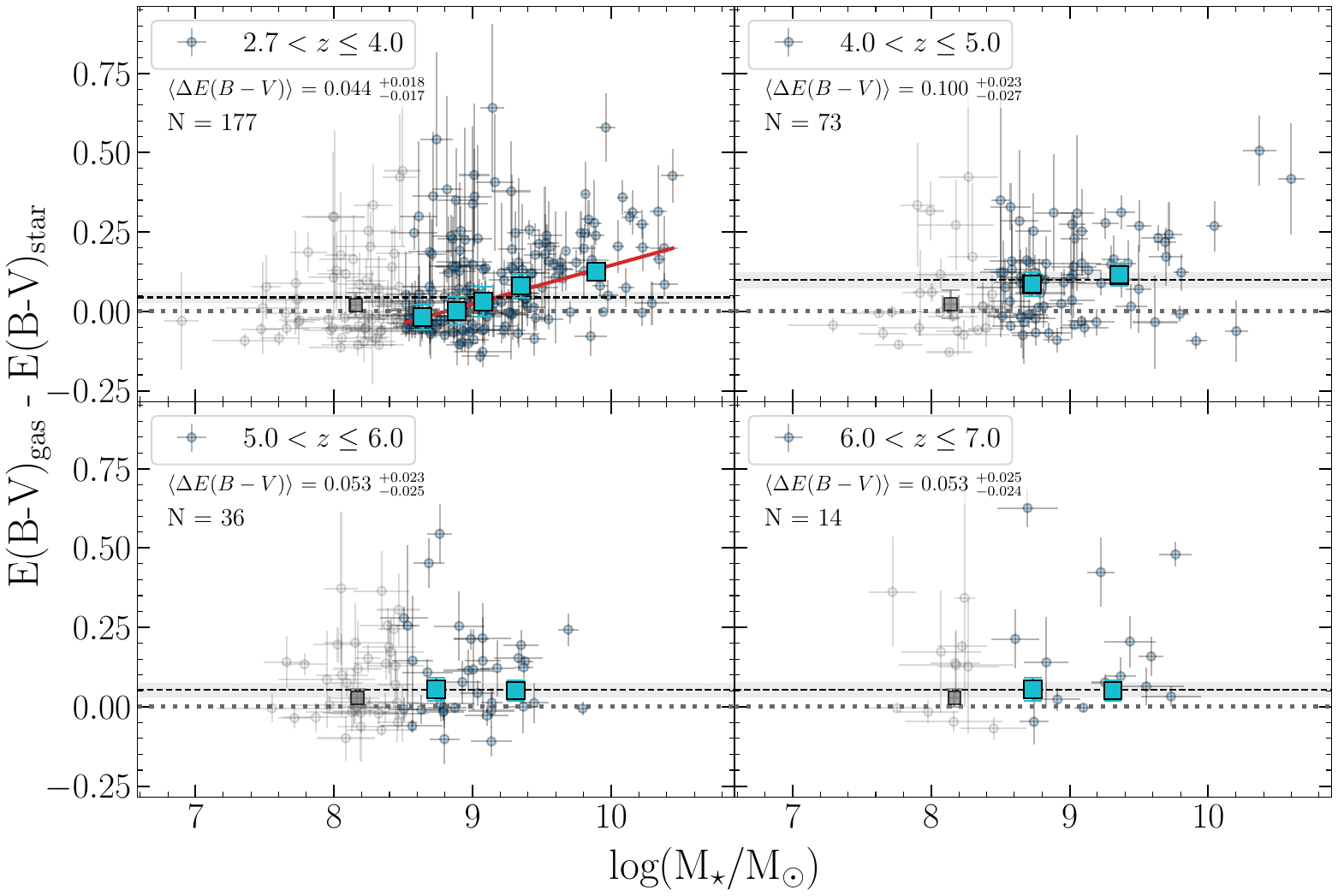}
    \caption{\ebvgas{}$-$\ebvstar{} versus stellar mass in bins of redshift. Individual galaxies are plotted as blue circles; stacked spectra are shown as cyan squares. We plot measurements from the combined $5.0 \leq z < 7.0$ composite spectra in both the $5.0 < z \leq 6.0$ and $6.0 < z \leq 7.0$ redshift bins. Greyed-out points indicate measurements that are not in the mass-representative range of our sample. The horizontal dashed line shows the median vertical offset computed from the stacks, with the $1\sigma$ uncertainty shaded in grey. $\langle\Delta E(B\!-\!V)\rangle$ reports the mean offset and uncertainty. The grey dotted line indicates where \deltaebv{} $ = 0$. In the $2.7 < z \leq 4.0$ bin, we plot the best fit to the stacked spectra as a solid red line.}
    \label{fig:4}
\end{figure*}

\begin{figure*}[ht!]
    \centering
    \includegraphics[width=0.95\linewidth]{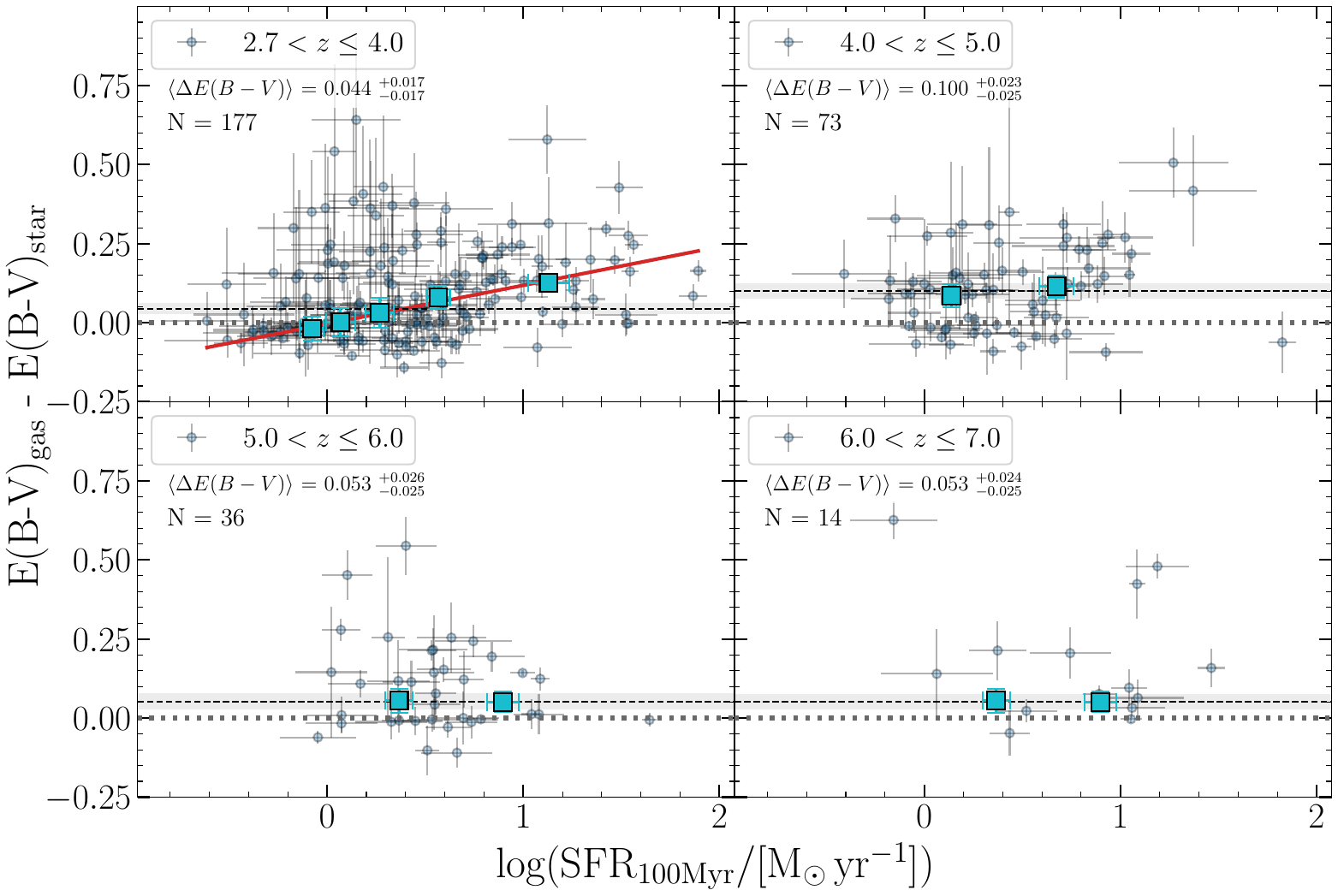}
    \caption{\ebvgas{}$-$\ebvstar{} versus $\mathrm{SFR}_{100 \rm Myr}$ in bins of redshift. Individual galaxies are plotted as blue circles; stacked spectra are shown as cyan squares. We plot measurements from the combined $5.0 \leq z < 7.0$ composite spectra in both the $5.0 < z \leq 6.0$ and $6.0 < z \leq 7.0$ redshift bins. The horizontal dashed line shows the median vertical offset computed from the stacks, with the $1\sigma$ uncertainty shaded in grey. $\langle\Delta E(B\!-\!V)\rangle$ reports the mean offset and uncertainty. The grey dotted line indicates where \deltaebv{} $ = 0$. In the $2.7 < z \leq 4.0$ bin, we plot the best fit to the stacked spectra as a solid red line.}
    \label{fig:5}
\end{figure*}

Due to the fact that many high-redshift galaxies lack multiple strongly detected Balmer lines, direct measurements of nebular reddening are often unobtainable. Meanwhile, measurements of stellar reddening are more frequently available from SED fitting. To quantify how stellar reddening can be translated into nebular reddening using our sample, we examine the differential reddening, $\Delta E(B\!-\!V) \equiv E(B\!-\!V)_{\rm gas} \: - E(B\!-\!V)_{\rm star}$, as a function of stellar mass and \sfrone{} (i.e., \logsfr{}). Predictions based on locally calibrated stellar-to-nebular conversions (e.g., \citealt{Calzetti+00}) are known to differ from observed nebular reddening in typical high-redshift star-forming galaxies (\citealt{Price+14}; \citealt{Reddy+15}; \citealt{Shivaei+20}), motivating an empirical recalibration at $z \gtrsim 3$ that reflects an explicit dependence on inferable galaxy properties.

Figures \ref{fig:4} and \ref{fig:5} show $\Delta E(B\!-\!V)$ versus stellar mass and \sfrone{}, respectively, for each redshift bin. The reported mean differential reddening values, $\langle\Delta E(B\!-\!V)\rangle$, are calculated from the stacked spectra.  We perturb the mean differential reddening by its constituent errors 1000 times, adopting the standard deviation of the mean as the corresponding $1\sigma$ uncertainty. Across the full redshift range, we also compute the mean differential reddening of individual galaxies within each redshift bin, finding a typical value of $\sim 0.1$ (i.e., additional reddening toward nebular regions than towards the stellar continuum). However, there is substantial scatter in the individual galaxies, with a typical $\sigma \sim 0.14$. This mean offset differs from that of the stacked spectra, which yield lower $\langle\Delta E(B\!-\!V)\rangle$ values ($\sim$ 0.06 on average). We note that the stacks provide higher S/N constraints on the typical Balmer decrement, are less sensitive to scatter and outliers, and comprise a larger effective sample as \hbeta{} detections are not required. Therefore, we adopt the stacked measurements as our fiducial basis for a practical stellar-to-nebular reddening conversion.

In the lowest-redshift bin ($2.7 < z \leq 4.0$), $\Delta E(B\!-\!V)$ shows evidence for dependence on both $M_\star$ and \sfrone{}. We compute the Bayesian information criterion (BIC) for both zero-slope (flat) and non-zero slope (linear) models, finding $\Delta \rm BIC \approx -17$. This value indicates that a linear dependence is strongly favored over a mass- or SFR-independent $\Delta E(B\!-\!V)$ in this redshift bin. Weighted least-squares fits to the stacked spectra yield slopes of $m_{\rm log M_\star} = 0.121 \pm 0.011$ mag per dex, and $m_{\rm SFR_{100}} = 0.122 \pm 0.011$ mag per dex. Although the inferred trend is shallow ($\Delta E(B-V)$ varies by at most $\sim 0.25$ over the full $M_\star$ and \sfrone{} range), the positive slope indicates systematically higher reddening of nebular emission relative to the stellar continuum in higher-mass and higher-\sfrone{} galaxies.

At $z \gtrsim 4$, our constraints on any dependence of $\Delta E(B\!-\!V)$ on $M_\star$ or \sfrone{} are limited by the small number of stacks. Within each redshift bin we have two stacked measurements, and in each case the resulting $\Delta E(B-V)$ values are consistent with each other within $1\sigma$. We therefore find no evidence in the stacked data for a mass- or \sfrone{}-dependent $\Delta E(B-V)$ at these redshifts and adopt a constant $\langle\Delta E(B\!-\!V)\rangle$ in each bin. The inferred mean offsets are positive, with  $\langle\Delta E(B\!-\!V)\rangle = 0.100^{+0.027}_{-0.026}$ at $4.0 \leq z < 5.0$, and $\langle\Delta E(B\!-\!V)\rangle = 0.053^
{+0.025}_{-0.025}$ at $5.0 \leq z < 7.0$, suggestive of a smaller differential reddening above $z \sim 5$. 

Here, we note the $4.0 \leq z < 5.0$ subsample is known to have an anomalous lack of UV-bright targets at $\rm log(M_\star/M_\odot) \gtrsim 10$ (\citealt{Clarke+25}), as previously discussed in Section \ref{subsec:sample}. If relatively unobscured galaxies are preferentially missed, then the stack at fixed stellar mass will be biased toward galaxies with higher \ebvgas{}, thereby biasing the mean \deltaebv{} high in this redshift bin. Additionally, because our $z > 4$ samples only sparsely populate the high-mass regime probed at $z \sim 2.7-4$, they may be unable to capture a trend if it arises primarily at the massive end.

For the practical purpose of estimating nebular reddening from SED-derived properties, we provide redshift-dependent prescriptions calibrated on the composite spectra in this work and valid over the stellar mass and \sfrone{} range probed by our sample. At $4 \leq z < 7$, the stacked measurements within each redshift bin yield consistent $\Delta E(B-V)$ values across $M_\star$ and \sfrone{}, with no measurable dependence at the current level of precision. We therefore recommend adopting a constant offset in these bins. At $2.7 \leq z < 4.0$, where a shallow but significant trend with mass and \sfrone{} is detected, we provide linear relations based on fits to the stacked spectra.

At $2.7 \leq z < 4.0$, \deltaebv{} depends on stellar mass as follows:

\begin{equation}
\begin{aligned}
E(B\!-\!V)_{\rm gas} 
&= E(B\!-\!V)_{\rm star} + 0.044 \\ 
&\quad + 0.121 \times \bigl[\log(\mathrm{M}_\star/\mathrm{M}_\odot)-9.0\bigr], 
\end{aligned}
\end{equation}

and on \sfrone{} as:
\begin{equation}
\begin{aligned}
E(B\!-\!V)_{\rm gas}
&= E(B\!-\!V)_{\rm star} + 0.044 \\ 
& + 0.122 \times \bigl[\log (\mathrm{SFR}_{100\, \mathrm{Myr}}/[\mathrm{M}_\odot \;\mathrm{yr}^{-1}])\bigr] 
\end{aligned}
\end{equation}
Either parametrization may be adopted, depending on which of stellar mass or \sfrone{} is available. 

At $4.0 \leq z < 5.0$:
\begin{equation}
E(B\!-\!V)_{\rm gas} = E(B\!-\!V)_{\rm star} + 0.100
\end{equation}

At $5.0 \leq z < 7.0$:
\begin{equation}
E(B\!-\!V)_{\rm gas} = E(B\!-\!V)_{\rm star} + 0.053
\end{equation}

These prescriptions are intended for statistical applications and should not be extrapolated beyond the mass, \sfrone{}, or redshift ranges probed by our sample. Additionally, in Section \ref{subsec:limitations}, we quantify how much the attenuation curve used infer \ebvstar{} affects the conversion prescription for each redshift range.

\subsection{Reddening versus Metallicity}

\begin{figure*}[ht!]
    \centering
    \includegraphics[width=0.95\linewidth]{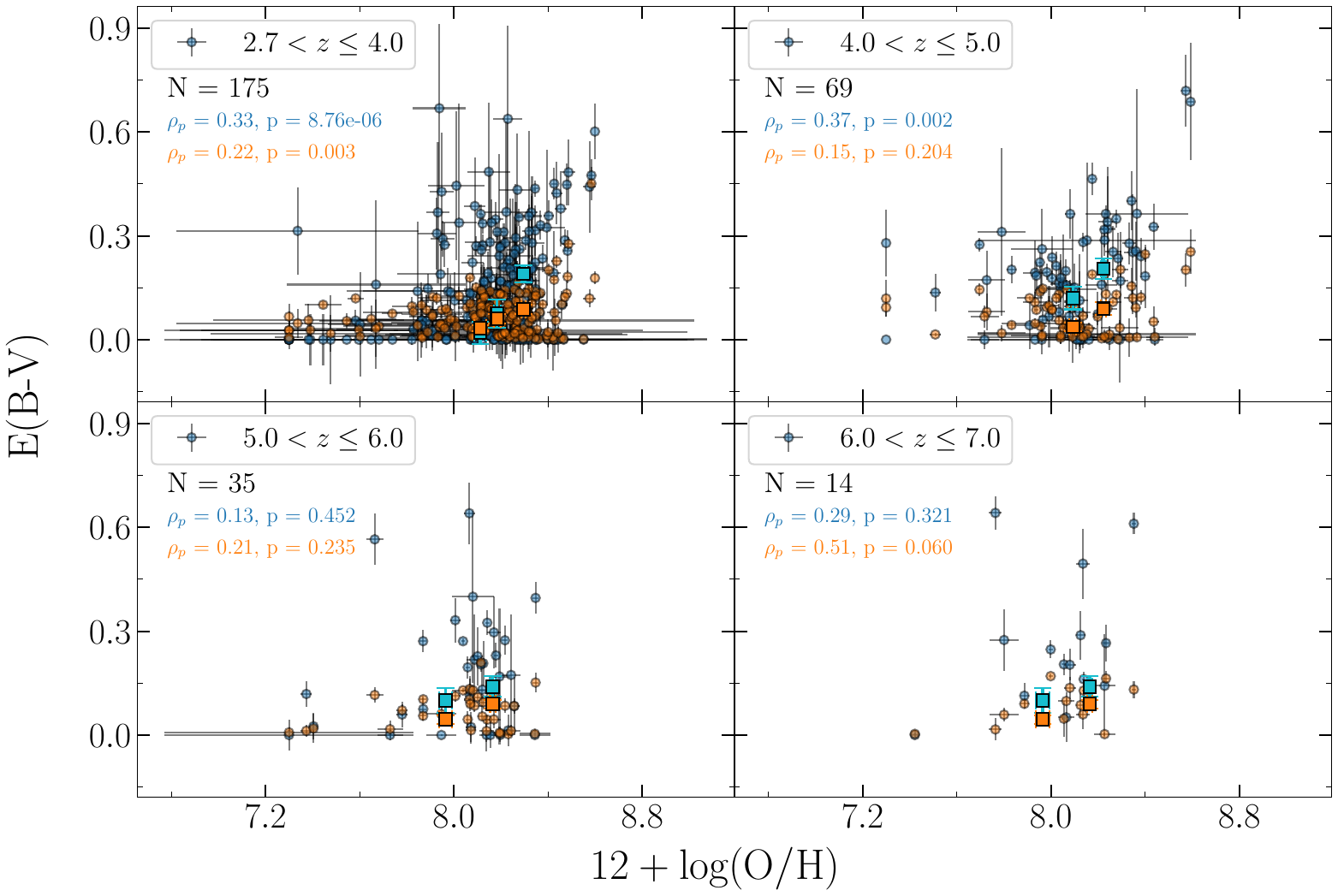}
    \caption{$E(B-V)$ versus metallicity, in bins of redshift. Blue points indicate \ebvgas{}, while orange points are \ebvstar{}. Stacks are shown as squares with the corresponding colors. We report the Pearson correlation coefficient, $\rho$, and $p$-value for the individual galaxies in each redshift bin.}
    \label{fig:6}
\end{figure*}

We next investigate reddening trends as a function of chemical enrichment by comparing nebular and stellar color excesses to gas-phase metallicity, $12 + \rm log(\mathrm{O/H})$. Figure \ref{fig:6} shows $E(B\!-\!V)_{\rm gas}$ (blue) and $E(B\!-\!V)_{\rm star}$ (orange) versus gas-phase metallicity, with stacked measurements overplotted as squares. For the individual galaxies in the two lowest redshift subsamples, nebular reddening and metallicity are positively correlated at the 3$\sigma$-level. Using a Pearson correlation test, we find $\rho = 0.33$ ($p = 8.76 \times 10^{-6}$) at $2.7 < z \leq 4.0$, and $\rho = 0.37$ ($p=0.002$) at $4.0 < z \leq 5.0$. In both bins, the stacked points follow the same qualitative increase of $E(B-V)_{\rm gas}$ towards higher $12+\log(\mathrm{O/H})$. In contrast, the $E(B\!-\!V)_{\rm star}$-metallicity relation for individual galaxies is weaker. At $2.7 < z \leq 4.0$, we find a modest correlation with 2$\sigma$ significance ($\rho = 0.21$, $p=0.003$), while at $z > 4.0$, there is no significant evidence for a correlation between \ebvstar{} and metallicity.

Overall, we find that gas-phase metallicity is more tightly correlated with $E(B\!-\!V)_{\rm gas}$ than with $E(B\!-\!V)_{\rm star}$ over $2.7 \leq z \leq 5$, which suggests that nebular regions are more closely influenced by dust associated with chemically enriched gas, whereas the reddening of the stellar continuum is less directly coupled to local metal abundance. At $z > 5$, neither relation is well-constrained with current sample sizes.

\section{Discussion} \label{sec:discussion}

Using JADES JWST/NIRSpec spectroscopy, including both individual galaxies and stacked spectra, we measure Balmer decrements for a large sample of galaxies spanning $2.7 < z < 7$. This approach enables a direct comparison between nebular and stellar reddening as a function of stellar mass, \sfrone{}, and gas-phase metallicity. Several trends emerge from this analysis. First, the Balmer decrement-$M_\star$ relation shows no significant evolution out to $z \sim 7$. Second, the typical differential reddening between nebular emission and stellar continuum emission is small but non-zero across $2.7 < z < 7$, with a linear dependence on mass and \sfrone{} at $z \sim 2.7-4$. We find no significant evidence for a dependence on mass or \sfrone{} at $z \gtrsim 4$, and indications that the mean offset decreases in magnitude towards higher redshift. Finally, we find that nebular reddening exhibits a clear correlation with gas-phase metallicity at $z \sim 3-5$, whereas the corresponding correlation for stellar reddening is either weaker or absent. Above $z = 5$, we do not find conclusive evidence for a significant correlation between either nebular or stellar reddening and metallicity with our current sample sizes.

Together, these results provide constraints on how dust content and geometry evolve in high-redshift galaxies, and on the extent to which stellar-to-nebular dust conversions calibrated at lower redshifts remain applicable in the early universe. In this section, we place these results in the context of previous work and discuss their physical implications.

\subsection{Balmer Decrement versus Stellar Mass Relation} \label{subsec:nebdisc}

At fixed $M_\star$, we do not detect significant redshift evolution in the Balmer decrement from $z \sim 3$ to $z \sim 7$, extending the non-evolution that was previously established out to $z \sim 2$ (\citealt{Dominguez+13}; \citealt{Kashino+13}; \citealt{Price+14}, \citealt{Shapley+22}). Our sample pushes these constraints to higher redshift and lower typical masses and SFRs than pre-JWST Balmer-decrement studies. This result is consistent with recent evidence based on data from the JWST CEERS and JADES surveys for a roughly constant Balmer decrement-$M_\star$ relation to $z \sim 7$ (\citealt{Shapley+23}; \citealt{Sandles+23}; \citealt{Woodrum+25}). Our analysis extends this work by incorporating stacked spectra and redshift binning, enabling a more robust test of evolution in the relation over $z \sim 3-7$. 

The Balmer decrement is sensitive to the dust column density distribution along sightlines to H II  regions, and therefore reflects both the total dust content and its spatial distribution. A lack of evolution in the Balmer decrement at fixed $M_\star$ is naturally explained if the effective dust column towards nebular regions (or equivalently, the dust mass surface density) primarily scales with the galaxy's assembled stellar mass, as previously suggested by \cite{Shapley+22}. Our results provide additional empirical support for this picture by showing that this relation holds out to $z \sim 7$. 

Here, we note that the ratio between dust mass and stellar mass ($M_{\rm dust}/M_\star$) in star-forming galaxies is known to evolve with redshift (\citealt{Shivaei+22}; \citealt{Jolly+25}). However, our interpretation does not require the dust mass at fixed stellar mass to be constant with redshift. Changes in characteristic scale or geometry, such as a clumpier dust distribution (\citealt{Witt+00}; \citealt{Seon+16}), will yield lower attenuation for the same dust mass. Similarly, if the dust covering fraction is less than unity, the Balmer decrement may also be reduced for a given total dust content (Reddy et al., in prep.).

\subsection{Comparison of Nebular and Stellar Reddening}
\label{subsec:nebstel}

We find the mean differential reddening $\Delta E(B-V) \equiv E(B-V)_{\rm gas} - E(B-V)_{\rm star}$ to be slightly positive but consistent with zero within the uncertainties in most redshift bins. In our lowest redshift bin, $\Delta E(B-V)$ increases with stellar mass and \sfrone{}, in that higher-mass and higher-SFR galaxies tend to show larger differential reddening. This trend is consistent with results from the MOSDEF survey at $z \sim 2$ (\citealt{Reddy+15}; \citealt{Sanders+21}; \citealt{Lorenz+23}). Quantitatively, our best-fit slope of $0.122 \pm 0.011$ for $\Delta E(B-V)$ versus \sfrone{} is shallower than the MOSDEF-based calibration in \citet{Sanders+21}, which reported a slope of $\sim0.54$. This difference is likely due to the differing redshift ranges probed, as well as our sample extending to lower typical SFRs, where both \ebvgas{} and \ebvstar{} are closer to zero (see Figure \ref{fig:5}).

In the local universe, the ratio between nebular and stellar reddening is around $\sim 2.27$ (\citealt{Calzetti+00}; hereafter C00). Subsequent studies have established that nebular reddening is typically higher than stellar reddening out to $z \sim 2$ (\citealt{Kashino+13}; \citealt{Price+14}; \citealt{Reddy+15}), but is very sensitive to the reddening curve assumed for the stellar continuum (e.g., \citealt{Reddy+20,Shivaei+20}). The observed difference between \ebvgas{} and \ebvstar{} is commonly interpreted using a two-component picture of attenuation (e.g., \citealt{Charlot+00}; \citealt{Price+14}). In this model, the first dust component arises from the optically thick dust in stellar birth clouds that primarily reddens emission lines from star-forming H II regions. This component is traced by \ebvgas{} and is associated with young stellar populations. The second component reflects the reddening of the stellar continuum by diffuse dust in the ISM. This component is traced by \ebvstar{} and affects the older, more spatially extended stellar populations that dominate the stellar continuum. Since the two components probe different dust columns, the framework naturally explains the systematic offset between \ebvgas{} and \ebvstar{} observed in the local  and $z < 2$ Universe.

\begin{figure}[ht!]
    \centering
    \includegraphics[width=0.48\textwidth]{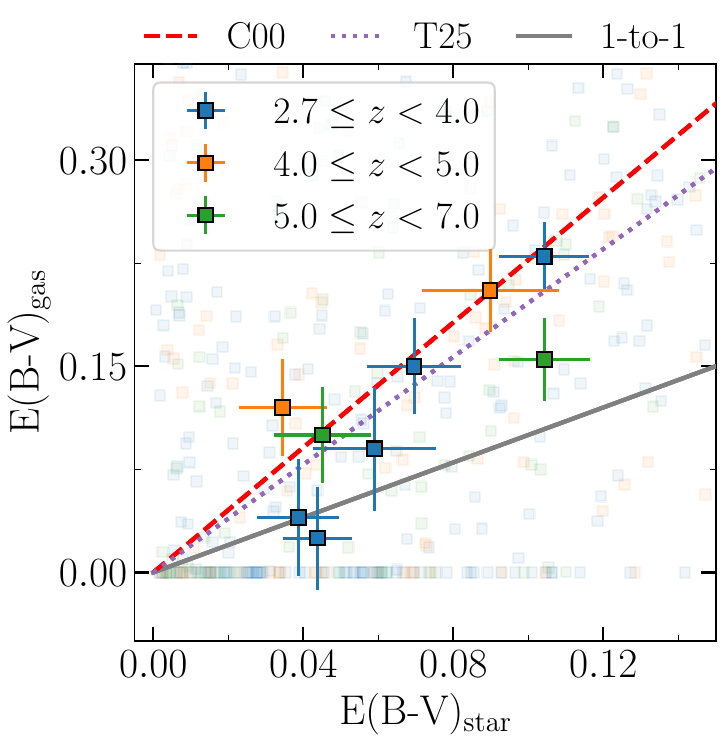}
    \caption{Comparison of \ebvgas{} and \ebvstar{} from stacked spectra. The low-opacity squares plot individual galaxies. The red dashed line indicates the \citet{Calzetti+00} relation; the purple dashed line the \citet{Tsujita+25} relation. The solid grey line is the one-to-one relation.}
    \label{fig:7}
\end{figure}

However, the nebular-to-stellar reddening ratio at higher redshifts is often lower than the local C00 value. Recent work from the ALPINE-CRISTAL-JWST survey at $z = 4-6$ \citep{Tsujita+25} (hereafter T25) reports \ebvgas{}$/$\ebvstar{}$ = 1.96^{+0.12}_{-0.14}$. In Figure \ref{fig:7}, we show \ebvgas{} versus \ebvstar{} for our stacked spectra, along with the local C00 relation (red dashed), the high-$z$ T25 relation (purple dotted), and the 1-to-1 line (solid grey). The majority of our stacked measurements are consistent with both the C00 and T25 relations within uncertainties.

\begin{figure*}[ht!]
    \centering
    \includegraphics[width=0.95\linewidth]{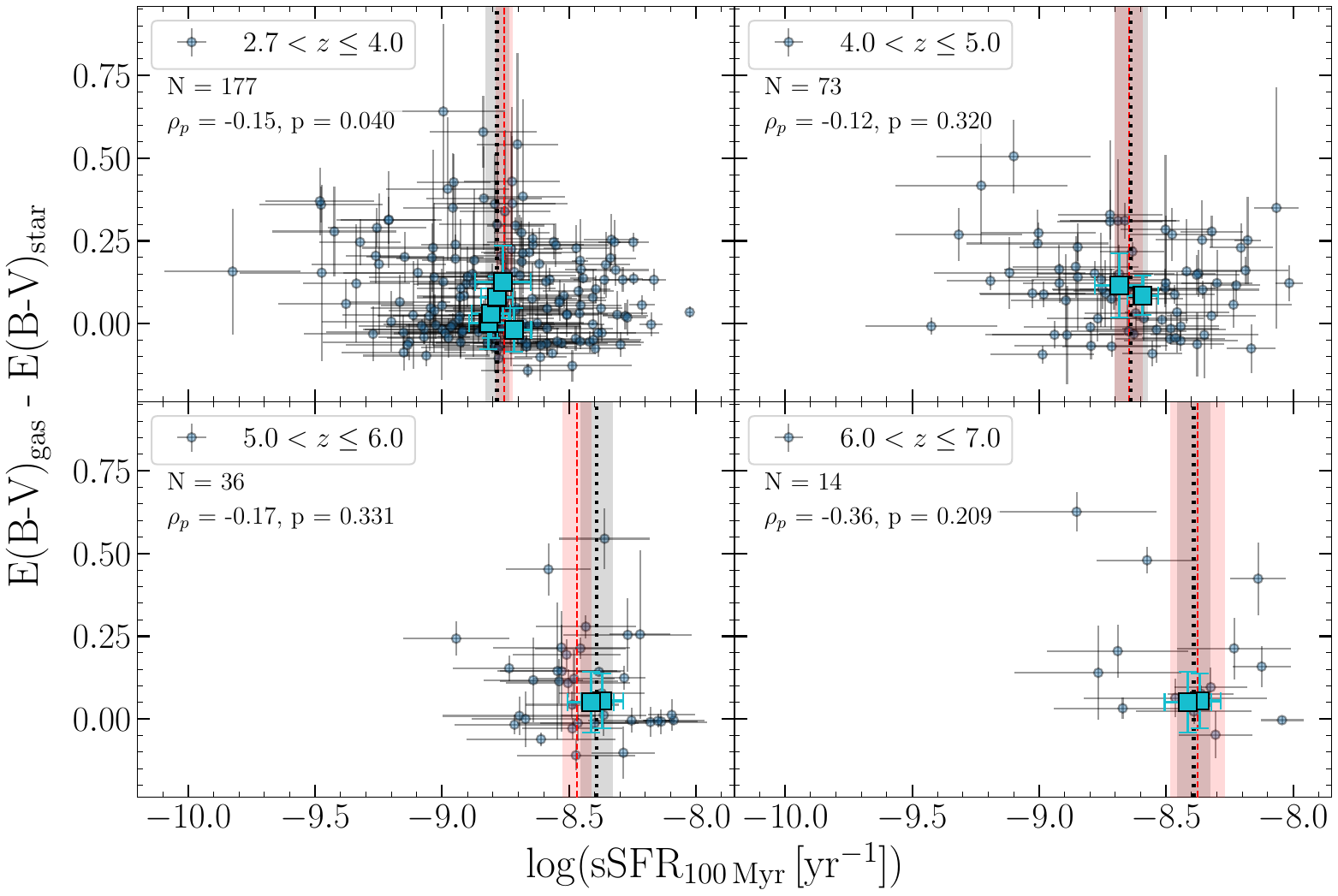}
    \caption{\ebvgas{}$-$\ebvstar{} versus $\mathrm{sSFR}_{100 \rm Myr}$ in bins of redshift. Individual galaxies are plotted as blue circles; stacked spectra are shown as cyan squares. We plot measurements from the combined $5.0 \leq z < 7.0$ composite spectra in both the $5.0 < z \leq 6.0$ and $6.0 < z \leq 7.0$ redshift bins. The dashed red line shows the median sSFR of the individual points; the black dotted line shows the median sSFR of the stacked points. The red and black shaded regions show the uncertainty on the median for the individual and stacked points, respectively. We report the Pearson correlation coefficient, $\rho$, and $p$-value for the individual galaxies in each redshift bin.}
    \label{fig:8}
\end{figure*}

Here, we note that two of the low-mass stacks in the $2.7 \leq z < 4.0$ bin lie near the one-to-one line, rather than the expected C00 or T25 relations. This deviation may be because $\sim43$\% of the individual galaxies contributing to these stacks have Balmer decrements either below or within 1$\sigma$ of the Case B recombination ratio (see Figure \ref{fig:2}), leading to low inferred stack \ebvgas{}. We discuss this caveat further in Section \ref{subsec:limitations}.

A notable outlier is the highest-mass stack at $5.0 \leq z < 7.0$, which lies below both the Tsujita+25 and Calzetti+00 trends. This disagreement is suggestive of a reduced contrast between the reddening of nebular regions and the stellar continuum at $z \gtrsim 5$, implying that nebular lines and the UV/optical continuum are probing similar effective dust columns.

A plausible physical interpretation is that the classic two-component attenuation picture weakens at early times. Locally, \ebvgas{} exceeds \ebvstar{} because nebular emission originates in compact H II regions still embedded in dusty birth clouds, whereas the continuum is dominated by light from stars that have moved into the more diffuse ISM (e.g., \citealt{Charlot+00}). At $z \gtrsim 5$, several effects could contribute to reducing this offset. One possibility is that the relative importance of diffuse ISM dust decreases at earlier times, and dust associated with star-forming regions regulates both stellar and nebular reddening, a scenario also discussed by \citealt{Reddy+20}. Consistent with this picture, \citealt{Lorenz+23} found at $z \sim 2$ that stellar attenuation does not correlate with inclination, contrary to expectations if diffuse ISM dust is a dominant component (as edge-on sightlines should increase the diffuse dust column). This result is interpreted with a three-component model: dust in star-forming regions that modulates stellar light, dust in $\sim$kpc-scale clumps that modulates nebular emission, and diffuse dust that is not significant for either stellar or nebular attenuation. In this model, nebular emission and continuum photons encounter similar dust columns. 

A further consideration is that if the rest-UV/optical continuum at these redshifts is increasingly dominated by young, high-sSFR populations, the effective separation between nebular and continuum sightlines may be reduced (\citealt{Wild+11,Price+14,Reddy+20,Maheson+25}). In Figure \ref{fig:8}, we plot \deltaebv{} as a function of $\log(\mathrm{sSFR_{100 \, Myr}})$ (hereafter s\sfrone{}) in each redshift bin. We quantify the trend using a Pearson correlation test on the individual galaxies, since the stacked points are too few and span a limited range in s\sfrone{}. We find a negative correlation at the $2\sigma$-level in the $2.7 < z \leq 4.0$ bin, in qualitative agreement with the picture in \citet{Price+14}. While \citet{Reddy+15} reported the opposite trend at $z \sim 2$, a direct comparison is complicated by differences in the adopted stellar reddening curve and sampled sSFR range. There is no apparent correlation in the higher-redshift bins. We also note that the median s\sfrone{} (denoted by vertical lines) in the $5.0 < z \leq 7.0$ redshift range is higher than in the lower-redshift bins, consistent with JADES-based results that the median s\sfrone{} of star-forming galaxies increases from $z \sim 3$ to $z \sim 7$ \citep{Sandles+22,Simmonds+25,Clarke+25}. Taken together, these trends support a picture in which higher-sSFR systems have more similar dust columns along nebular and continuum sightlines, driving the typical \ebvgas{}/\ebvstar{} towards unity.

A way to test whether nebular and stellar reddening are governed by similar dust geometries at $z \gtrsim 5$ would be to look for inclination effects using the axis ratio. In the two-component picture, \ebvstar{} should show a stronger dependence on inclination because the stellar continuum has a larger spatial extent and therefore longer effective path length through the ISM in an edge-on configuration. In contrast, \ebvgas{} should be comparatively less sensitive to inclination if nebular attenuation arises primarily from compact regions. If instead attenuation is governed by patchy star-forming regions, both \ebvstar{} and \ebvgas{} may show little dependence on inclination. Measuring the strength of the attenuation trends with axis ratio will distinguish between these scenarios.

\subsection{Connection to Metallicity}

Across $2.7 < z \leq 7$, we find that the nebular reddening $E(B-V)_{\rm gas}$ shows a clearer positive association than $E(B-V)_{\rm star}$ with gas-phase metallicity. The strongest evidence for this connection between \ebvgas{} and metallicity is observed at $2.7 < z \leq 5.0$, with weaker constraints at $z > 5$ due to small sample sizes and larger uncertainties. This result differs from MOSDEF results at $z \lesssim 2.6$ \citep{Shivaei+20}, in which stellar reddening is more tightly correlated than nebular reddening to metallicity. There are a few explanations for this discrepancy: differences in the redshift and mass or SFR ranges probed by the samples, different SED-fitting assumptions used to infer $E(B-V)_{\rm star}$, and differing strong-line calibrations used to derive metallicities. Such differences can affect both the dynamic range and apparent slope of reddening-metallicity trends. Given the higher S/N and greater sensitivity of JWST spectroscopy relative to ground-based surveys, our measurements should provide a more robust view of reddening-metallicity trends at $z \gtrsim 3$.

A simple physical picture for our result is that the dust modulating Balmer decrements is closely tied to enriched gas in H II regions. At fixed $M_\star$, more metal-rich galaxies have higher dust-to-gas ratios (\citealt{Li+19}) and therefore build up larger effective dust columns along H II-region sightlines (e.g., \citealt{Reddy+10}; \citealt{Dominguez+13}), boosting $E(B-V)_{\rm gas}$. In contrast, $E(B-V)_{\rm star}$ averages over a broader range of stellar ages and sightlines, and is thus less directly tied to gas-phase metallicity. These trends together suggest a picture in which chemical enrichment primarily modulates the extra reddening affecting nebular regions at $z \sim 3-5$, and does not affect reddening of the stellar continuum to the same degree.

This interpretation is not in tension with the reduced nebular-stellar offset discussed in Section \ref{subsec:nebstel}. The picture presented in the previous section is driven by the $z > 5$ stacks, where we do not find a significant correlation between \ebvgas{} and metallicity. By contrast, the \ebvgas{}-metallicity correlation is clearest at $z < 5$, where we also find a significantly non-zero offset between \ebvgas{} and \ebvstar{}, particularly at the highest masses. However, the physical connection between the dominant trends in these two different redshift ranges is still unclear. Larger samples at $z > 5$ will provide tighter constraints on potential redshift evolution in reddening-metallicity trends.

In the context of Section \ref{subsec:nebdisc}, the simplest overall interpretation is that $M_\star$ sets the typical effective dust column toward star-forming regions. The lack of evolution in the Balmer decrement-$M_\star$ relation thus implies that changes in the factors that control nebular reddening---dust-to-gas ratio, dust production, and dust geometry---must combine to produce a similar effective dust column at fixed mass across $z \sim 3-7$.

\subsection{Limitations and Caveats}
\label{subsec:limitations}

There are several assumptions and selection effects that limit how precisely we can interpret nebular and stellar dust reddening trends in our sample. One important assumption is the attenuation curve used to infer \ebvstar{} from the best-fit \textsc{Prospector} stellar population model. In practice, galaxies have attenuation curves that vary with dust composition and geometry, which can shift the inferred \ebvstar{} (e.g., \citealt{Theios+19}; \citealt{Reddy+20}; \citealt{Shivaei+20}). As described in Section \ref{subsubsec:sed}, we adopted either an SMC or Calzetti curve as the fiducial solution for each galaxy in our sample, motivated by previous studies (\citealt{Sanders+15,Xinnan+18,Clarke+25}). We report the number of galaxies assigned each curve solution across redshift bins in Section \ref{subsubsec:sed}. The large majority of galaxies in our sample are best fit with an SMC attenuation curve, whereas Calzetti-curve solutions appear mainly at the highest stellar masses, consistent with previous findings (\citealt{Reddy+18,Xinnan+18}). This distribution is also broadly consistent with other high-redshift studies that favor steeper attenuation curves, although alternative prescriptions to the SMC curve (such as MOSDEF-based attenuation curves) have also been used (\citealt{Reddy+20,Shivaei+20}).

To test the sensitivity of our results to this choice in attenuation curve, we compare our fiducial \ebvstar{} values to those obtained by instead assuming that all galaxies follow either an SMC or Calzetti law. We define $\delta E(B-V)_{\rm star, \, curve} \equiv E(B-V)_{\rm star, \, curve} - E(B-V)_{\rm star, \, fiducial}$, where ``curve" denotes the SMC-only or Calzetti-only case, and ``fiducial" refers to the values used in our analysis. Under an SMC-only prescription, the median $\delta E(B-V)_{\rm star, \, SMC}$ in each redshift bin is consistent with zero, indicating that our main trends are not strongly affected by adopting SMC for the full galaxy sample. An all-Calzetti assumption produces larger $\delta E(B-V)_{\rm star, \, Calzetti}$; however, because SMC-like attenuation is generally better motivated at high redshift (\citealt{Reddy+18}), we focus on the SMC-only case in the remainder of this discussion. For the stacks, $\delta E(B-V)_{\rm star, \, SMC}$ is consistent with zero within uncertainties for all but the highest-mass $2.7 \leq z < 4.0$ stack, which is also the stack with the largest fraction (36\%) of galaxies assigned a fiducial Calzetti curve. We also test the effect of this assumption on \ebvstar{} to \ebvgas{} conversion prescriptions (see Section \ref{subsec:conversion}). In the $4.0 \leq z < 5.0$ and $5.0 \leq z < 7.0$ bins, the mean differential reddening under an SMC-only assumption is consistent with the fiducial SMC+Calzetti values within uncertainties. In the $2.7 \leq z < 4.0$ bin, the fitted slope of the stack \deltaebv{} with stellar mass under the SMC-only assumption also remains consistent with the fiducial SMC+Calzetti results at the 1$\sigma$ level.

A second important assumption in this work concerns the intrinsic Balmer line ratios adopted to convert the observed Balmer decrements into nebular reddening estimates. The intrinsic Balmer decrement shows little variation across the range of physical conditions expected in star-forming regions (\citealt{Smith+22}; \citealt{Sandles+23}); the Balmer decrement changes by $\lesssim 10$\% with variations in $T_e \sim$ 5,000$-$20,000 K and $n_e \sim 10^3-10^4 \rm cm^{-3}$. However, recent work reports a prevalence of recombination line ratios that are unphysical under typical Case-B assumptions at high redshifts (e.g., \citealt{Yanagisawa+24,Cameron+24,Topping+24,Woodrum+25,McClymont+25}). In particular, \citet{Yanagisawa+24} explores the possibility that assuming Case B recombination may not be valid for all high-redshift star-forming galaxies. In our $2.7 < z \leq 4.0$ subsample, $\sim 6$\% ($\sim 13$\%) of points in the mass-representative range are below Case B at the 3$\sigma$ (1$\sigma$) level. For this small minority of galaxies in our sample, assuming an intrinsic Case B Balmer decrement may cause the inferred \ebvgas{} values to be underestimated. Additionally, as noted in Section \ref{subsec:sample}, the highest-mass stack in the $4.0 \leq z < 5.0$ bin appears to miss some UV-bright systems. This incompleteness could bias the reddening values inferred for that stack high, affecting our reported reddening conversion prescription.

These considerations are especially relevant for interpreting evolution at the highest redshifts. Future work can test the robustness of our conclusions by repeating the analysis under other known attenuation curves (e.g., MOSDEF; \citealt{Reddy+20}), using additional hydrogen recombination lines (e.g., Paschen lines; \citealt{Reddy+23a}; \citealt{Reddy+25}) to cross-check nebular reddening, and improving mass completeness in the $4.0 \leq z <5.0$ redshift range. In particular, deeper spectroscopy and a larger sample at $z \gtrsim 5$, especially at \logmass{}$ \gtrsim 9$, would better test our conclusions at the highest redshifts.

\section{Conclusions} \label{sec:conclusion}


With JWST/NIRSpec spectroscopy from the JADES survey, we measured nebular attenuation at $2.7 < z < 7.0$ from Balmer decrements for 293 individual galaxies and from stacked spectra incorporating 327 galaxies. We estimated stellar attenuation independently from SED fitting, enabling a direct comparison between nebular and stellar attenuation as a function of stellar mass, \sfrone{}, and gas-phase metallicity over the first $\sim$1-2 Gyr of cosmic time. We summarize our main findings below:


\begin{enumerate}
    \item \textit{The Balmer decrement-\mstar{} relation shows no significant evolution out to $z\sim7$.}
    At fixed stellar mass, the typical \halpha{}$/$\hbeta{} ratios are consistent across redshift bins within uncertainties, implying that the mass-attenuation relation is largely in place by $z\sim7$. This result suggests that the average dust columns at fixed \mstar{} do not evolve significantly over this timescale, despite changes in galaxy growth and ISM conditions.

    \item \textit{The typical differential reddening \deltaebv $\equiv$ \ebvgas{} $-$ \ebvstar{} across the full $z=2.7-7.0$ redshift range is modest (around $\sim 0.04$), with the strongest trend at $z \sim 3$ and evidence for decreasing contrast at higher redshift.}
    In the lowest-redshift bin, higher-mass and higher-\sfrone{} galaxies show larger \deltaebv, consistent with nebular regions probing higher effective dust columns than the stellar continuum. At $z\gtrsim4$, we do not find significant trends with mass or \sfrone{}; the inferred \deltaebv{} values are generally small and consistent with zero within 3$\sigma$ above $z \sim 5$. This result implies that \ebvgas{} and \ebvstar{} may reflect similar effective dust columns at higher redshifts. We additionally provide a prescription to convert the SED-derived \ebvstar{} and \logmass{} (or \logsfr{}) into an \ebvgas{} estimate when Balmer decrement measurements are unavailable.

    \item \textit{Nebular reddening correlates with gas-phase metallicity at $z\sim3-5$, while the stellar reddening-metallicity relation is weaker or absent.}
    At fixed \mstar{}, galaxies with higher $12 + \log(\rm O/H)$ exhibit larger \ebvgas{}, consistent with nebular attenuation being set by the effective dust column along H II region sightlines that grows with chemical enrichment. The comparatively weaker \ebvstar{}-metallicity dependence can be explained by the fact that continuum attenuation probes a mixture of sightlines through the diffuse ISM.
\end{enumerate}

Together, these results suggest that the effective dust column towards H II regions is already tightly correlated with assembled stellar mass by $z \sim 7$, and at $z < 5$, metallicity acts as a secondary dependence that modulates nebular sightlines more strongly than the integrated continuum below $z \sim 5$. The behavior of the \ebvgas{}-\ebvstar{} relation emphasizes that modeling and interpreting high-redshift dust requires accounting for not only dust mass, but dust geometry as well. Larger NIRSpec samples that better populate the high-mass $z \gtrsim 5$ regime will enable robust testing of these relations, and will sharpen constraints on how dust geometry and chemical enrichment regulate obscuration in the early universe.


\section*{Acknowledgements}
We would like to acknowledge the JADES team for their efforts in designing, executing, and making public their JWST/NIRSpec and JWST/NIRCam survey data. We also acknowledge support from NASA grants JWST-GO-01914 and JWST-GO-03833, and NSF AAG grants 2009313, 2009085, 2307622, and 2307623. This work is based on observations made with the NASA/ESA/CSA James Webb Space Telescope as well as the NASA/ESA Hubble Space Telescope. Data were obtained from the Mikulski Archive for Space Telescopes at the Space Telescope Science Institute, which is operated by the Association of Universities for Research in Astronomy, Inc., under NASA contract NAS5-03127 for JWST and NAS 5–26555 for HST. Data were also obtained from the DAWN JWST Archive maintained by the Cosmic Dawn Center. The specific observations analyzed can be accessed via \dataset[doi:10.17909/8tdj-8n28]{https://archive.stsci.edu/doi/resolve/resolve.html?doi=10.17909/8tdj-8n28}, \dataset[doi:10.17909/gdyc-7g80]{https://archive.stsci.edu/doi/resolve/resolve.html?doi=10.17909/gdyc-7g80}, and \dataset[doi:10.17909/fsc4-dt61]{https://archive.stsci.edu/doi/resolve/resolve.html?doi=10.17909/fsc4-dt61}. This work used computational and storage services associated with the Hoffman2 Cluster which is operated by the UCLA Office of Advanced Research Computing’s Research Technology Group.

\bibliography{bibliography}
\bibliographystyle{aasjournal}

\end{document}